\documentclass[sigconf]{acmart}

\usepackage{graphicx}
\usepackage{booktabs}
\usepackage{multirow}
\usepackage{amsmath}
\usepackage{amsfonts}
\usepackage{array}
\usepackage{pifont}
\usepackage{float}
\usepackage{xspace}
\usepackage{xcolor}
\usepackage{soul}

\copyrightyear{2026}
\acmYear{2026}
\setcopyright{cc}
\setcctype{by-nc-nd}
\acmConference[CIKM '26]{Proceedings of the 35th ACM International Conference on Information and Knowledge Management}{November 07--11, 2026}{Rome, Italy}
\acmBooktitle{Proceedings of the 35th ACM International Conference on Information and Knowledge Management (CIKM '26), November 07--11, 2026, Rome, Italy}
\acmDOI{10.1145/3799682.3841062}
\acmISBN{979-8-4007-2539-5/2026/11}
\newcommand{\rev}[1]{#1}
\newcommand{\eg}{\textit{e.g.}\xspace}

\begin{document}


\title{H2CE: Modeling Geo-Semantic Interactions for POI Reranking with Heterogeneous Two-Stage Cross-Encoders}

\author{Zhengwei Bai}
\affiliation{\institution{Amazon}\city{Sunnyvale}\country{USA}}
\email{zwbai@amazon.com}

\author{Moreno D'Inc\`{a}}
\authornote{Work performed during an internship at Amazon.}
\affiliation{\institution{University of Trento}\city{Trento}\country{Italy}}
\email{moreno.dinca@unitn.it}

\author{Danielle Class}
\affiliation{\institution{Amazon}\city{Boston}\country{USA}}
\email{dcclss@amazon.com}

\author{Alessandro Moschitti}
\affiliation{\institution{Amazon}\city{\rev{Los Angeles}}\country{USA}}
\email{amosch@amazon.com}

\begin{abstract}
Point-of-Interest (POI) reranking in local search must model query-conditioned tradeoffs among lexical semantics, geospatial proximity, and numerical quality signals such as rating and review count, while remaining practical under real-time serving constraints. \rev{A close POI may only partially satisfy the query intent, while a farther one may offer stronger semantic and quality evidence.} We present H2CE, a Heterogeneous Two-stage Cross-Encoder for latency-bounded POI reranking. H2CE represents numerical attributes in two complementary ways: bucketized natural-language descriptors are inserted into the cross-encoder input to support semantic--numeric attention, while exact scalar values are processed by dedicated MLPs to preserve magnitude information. The resulting semantic and numerical embeddings are fused through latent-space aggregation, enabling nonlinear interactions beyond scalar weighted sums. H2CE then applies a two-stage architecture: Stage~1 scores all candidates pointwise for scalable filtering, and Stage~2 performs head-to-head pairwise comparison among the top-$K$ candidates with Copeland aggregation, making fine-grained relative tradeoffs explicit while reducing pairwise cost from $O(N^2)$ to $O(N+K(K-1))$. On a 5,743-query local search test set, H2CE achieves 67.48\% NDCG@5, improving over XGBoost LTR by +22.82\% absolute and over a zero-shot LLM reranker by +35.89\%. The pairwise stage adds +1.98\% NDCG@5 over the pointwise model alone. Ablations confirm the value of numerical features, latent aggregation, top-$K$ pairwise reranking, and aligned training.
\end{abstract}
\vspace{-.5em}

\ccsdesc[500]{Information systems~Retrieval models and ranking}
\ccsdesc[300]{Information systems~Search engine architectures and scalability}
\ccsdesc[300]{Information systems~Learning to rank}
\vspace{-.5em}
\keywords{Point-of-Interest Reranking; Local Search; Learning to Rank; Heterogeneous Feature Fusion; Cross-Encoder; Pairwise Ranking}
\vspace{-.5em}

\maketitle

\vspace{-.5em}
\section{Introduction}
Local search has become a core capability of voice assistants, map services, and conversational agents: users routinely ask for \emph{``the closest pharmacy,''} \emph{``a top-rated Italian restaurant near me,''} \emph{``true Italian restaurants near the Eiffel Tower,''} or \emph{``Walgreens open 24 hours.''} The quality of Point-of-Interest (POI) ranking directly affects how often users find what they are looking for, and how often they re-engage with the product~\cite{psyllidis2022points,liu2017experimental,mei2023improving}. Because local search operates over millions of POIs under strict latency and cost constraints, modern systems decompose the problem into retrieval and reranking stages: retrieval narrows the search space, while reranking refines the final ordering over a smaller but still potentially large candidate set. Even at the reranking stage, the pool may contain hundreds or thousands of POIs, making expensive listwise or LLM-based reranking over all candidates impractical for real-time serving. A deployable reranker must therefore improve semantic and numerical reasoning while preserving bounded inference cost.

In this paper we focus on the reranking stage. Our starting point is a commercial local search pipeline that, like many systems in the industry, relied for years on query-type classification, type-specific retrievers, and an XGBoost~\cite{xgboost} reranker operating on hand-crafted features such as BM25-style~\cite{robertson2009probabilistic} textual scores, Jaccard/Levenshtein distances, geographic proximity, and engagement signals, together with heuristic chain rankers and blending rules. While efficient, this design suffers from two limitations that we found to be dominant drivers of quality loss.
 
POI relevance is not purely textual and not purely numerical. It depends on interactions among query semantics, POI semantics, user or target location, POI location, and numerical quality attributes such as distance, rating, and review count. Consider the query \emph{``true Italian restaurants near the Eiffel Tower.''} A very close candidate whose evidence suggests a mixed Spanish--Italian restaurant may be less desirable than a slightly farther POI whose name, categories, menu, and descriptions strongly indicate an authentic Italian restaurant. However, this preference is not absolute: if the authentic Italian restaurant is only a few meters farther away, semantic match should likely dominate; if it is more than a kilometer farther away, the closer mixed-cuisine restaurant may be preferable, especially when the query emphasizes proximity or walking distance. Thus, distance cannot be used as a trivial reranking feature independently of lexical semantics. Its contribution depends on query intent, POI evidence, and the magnitude of the geographic difference. We refer to this setting as \emph{geo-semantic reranking}: POI ranking where geographic proximity and numerical quality signals are interpreted jointly with query intent and POI semantics, rather than as independent ranking features.

The same issue appears across common local-search intents. For \emph{``closest pharmacy,''} distance should dominate once candidates satisfy the pharmacy intent. For \emph{``best Italian restaurant,''} ratings and review evidence should matter more, but only among candidates that semantically match the cuisine. For \emph{``Walgreens open 24 hours near me,''} entity match and operational constraints may dominate rating or popularity. \rev{Spatial and numerical features are thus not independent side features, which creates a modeling challenge for standard rerankers.} 

Transformer-based cross-encoders~\cite{nogueira2019passage,lin2021pretrained,chen2024bge,bge_embedding} excel at semantic matching between query and POI text, but raw numerical values such as \texttt{``0.3\,mi''}, \texttt{``4.7 stars''}, or \texttt{``823 reviews''} are poor semantic objects for attention. Conversely, gradient-boosted trees~\cite{xgboost} exploit structured numerical features precisely, but provide only shallow semantic matching. A common solution is to compute a neural semantic score and combine it with numerical feature scores through a weighted sum. However, this late scalar fusion cannot let the semantic encoder itself condition query--POI matching on coarse spatial and quality concepts, and it cannot easily express interactions such as distance mattering more for \emph{``near me''} or \emph{``walking distance''} queries, ratings mattering more for \emph{``best''} queries, or review count mattering more when reliability is implied.

We present H2CE (\textbf{H}eterogeneous \textbf{Two}-stage \textbf{C}ross-\textbf{E}ncoder), a serving-constrained POI reranker for query-conditioned semantic--numeric interaction. H2CE combines two complementary mechanisms. First, its \emph{Heterogeneous Feature Encoder} represents numerical attributes in two forms: bucketized natural-language descriptors inside the cross-encoder input, and exact scalar values processed by dedicated numerical MLPs. The semantic flow converts distance, rating, and review count into disambiguated descriptors such as \emph{``walkable,''} \emph{``near,''} \emph{``excellent,''} and \emph{``substantial,''} allowing transformer self-attention to condition semantic matching on coarse spatial and quality concepts. The numerical flow preserves fine-grained scalar precision lost by bucketization. H2CE then fuses the cross-encoder CLS embedding and numerical embeddings through latent-space aggregation, using a learned non-linear classifier rather than a scalar weighted sum. This enables interactions such as preferring a slightly farther POI when semantic match and review evidence are stronger, or preferring the closer POI when candidates are semantically and qualitatively similar.

Second, H2CE uses a two-stage pointwise-plus-pairwise architecture. Stage~1 applies the heterogeneous encoder pointwise to all retrieved candidates, producing a scalable coarse ranking and selecting the top~$K$. Stage~2 performs direct head-to-head comparison among those top-$K$ candidates using the same heterogeneous representation adapted to pairwise inputs. This makes relative feature tradeoffs explicit: the pairwise model can compare the semantic evidence, distance, rating, and review count of two competing POIs directly, and Copeland aggregation converts these preferences into the final top-$K$ ordering. Restricting pairwise reranking to $K{=}10$ keeps inference tractable, reducing the comparison cost from full $O(N^2)$ reranking to $O(N+K(K-1))$ ordered non-self comparisons.

We report NDCG~\cite{jarvelin2002ndcg} and Rank-Deviation Quality (RDQ)~\cite{zhou2026rankdeviation}. RDQ is designed for rankings where item quality is defined by relative preferences among candidates rather than by absolute category labels. On a 5,743-query local search test set, H2CE achieves 67.48\% NDCG@5 and 60.77\% RDQ@5 -- providing improvements of +22.82\% absolute NDCG@5 and +20.96\% absolute RDQ@5 improvement over an XGBoost learning-to-rank baseline and +35.89\% NDCG@5 over a zero-shot LLM reranker (RankGPT~\cite{sun2023chatgpt}) and +9.62\% over a BGE reranker model~\cite{bge_embedding} with local search fine-tuning. We further provide ablations of numerical features across stages, training strategies, top-$K$ sensitivity, quantization design, and serving latency.

In summary, this paper makes five contributions. We formulate POI reranking as query-conditioned semantic--numeric interaction, where geospatial and numerical attributes must be interpreted jointly with query intent and POI semantics. We introduce a Heterogeneous Feature Encoder that exposes numerical attributes both as bucketized descriptors inside the cross-encoder and as exact scalar values through numerical MLPs. We propose latent-space feature aggregation to capture conditional interactions that scalar score fusion cannot express. We introduce a two-stage pointwise-plus-pairwise reranking architecture that improves fine-grained ordering while bounding pairwise cost. Finally, we provide a comprehensive empirical study comparing H2CE against XGBoost LTR, zero-shot neural rerankers, and fine-tuned cross-encoder variants.

\begin{table*}[t]
    \centering
    \small
    \caption{
     Example candidate set for the query \emph{``true Italian restaurants near the Eiffel Tower.''}
    The table illustrates that semantic match, distance, rating, and review count can induce conflicting ranking preferences.
    }
     
    \label{tab:poi_reranking_example}
    \begin{tabular}{p{0.13\textwidth} p{0.30\textwidth} p{0.06\textwidth} p{0.06\textwidth} p{0.06\textwidth} p{0.27\textwidth}}
        \toprule
        \textbf{POI} & \textbf{Semantic evidence} & \textbf{Distance} & \textbf{Rating} & \textbf{Reviews} & \textbf{Ranking issue} \\
        \midrule
        \textbf{Roma Trattoria} 
        & Italian restaurant; pasta; trattoria; Roman cuisine 
        & 0.43 mi 
        & 4.6 
        & 820 
        & Strong semantic match; moderately close \\
        
        \textbf{Tapas \& Pasta} 
        & Spanish tapas; pasta; Italian dishes 
        & 0.07 mi 
        & 4.5 
        & 1,100 
        & Closest, but mixed cuisine \\
        
        \textbf{Italia Bella} 
        & Italian restaurant; pizza; pasta; wine bar 
        & 0.51 mi 
        & 4.7 
        & 390 
        & Strong match; slightly farther \\
        
        \textbf{Napoli Pizza} 
        & Pizza; casual Italian; takeaway 
        & 0.29 mi 
        & 4.1 
        & 95 
        & Close, but weaker quality evidence \\
        
        \textbf{Le Bistro Parisien} 
        & French bistro; brasserie; wine 
        & 0.05 mi 
        & 4.8 
        & 2,000 
        & Very close and popular, but wrong cuisine \\
        
        \textbf{Osteria Firenze} 
        & Authentic Italian; Tuscan cuisine; handmade pasta 
        & 1.32 mi 
        & 4.8 
        & 610 
        & Best semantic/quality match, but far \\
        \bottomrule
    \end{tabular}
\end{table*}


\section{Related Work}

\paragraph{POI search, recommendation, and geographic relevance.}
Research on POI search spans retrieval, recommendation, geographic modeling, and reranking~\cite{psyllidis2022points}. Classical POI recommendation combines collaborative filtering with geographic constraints or spatial regularization~\cite{ye2011exploiting,lian2018geomf}, while next-POI recommendation models exploit spatio-temporal context, trajectories, and mobility patterns~\cite{zhao2019stgn,luo2021stan,wang2020poi_context}. These methods establish the importance of geographic and contextual signals, but they mainly target personalized recommendation or trajectory prediction rather than query-dependent reranking over retrieved POI candidates.

More directly related, Query--POI matching models combine textual relevance with geographic or spatio-temporal context. Zhao et al.~\cite{zhao2019querypoi} incorporate semantic similarity with geographic correlation, and Yuan et al.~\cite{yuan2020stdgat} use spatio-temporal dual graph attention for Query--POI matching. These works recognize that POI relevance is not purely textual. H2CE differs in task and mechanism: it focuses on latency-constrained reranking after retrieval, and represents geospatial/numerical attributes both as bucketized natural-language descriptors inside the cross-encoder and as exact scalar values through numerical MLPs. This supports query-conditioned geo-numerical tradeoffs such as when proximity, cuisine match, rating, or review evidence should dominate.
 
\paragraph{Neural reranking and heterogeneous features.}
Cross-encoders jointly encode a query and candidate through self-attention and are among the strongest neural rerankers~\cite{nogueira2019passage,thakur2021beir,lin2021pretrained}, with common backbones including BERT~\cite{devlin2019bert}, BGE~\cite{chen2024bge}, and ModernBERT~\cite{modernbert2024}. \rev{TANDA~\cite{garg2020tanda} established the first transformer-based state of the art for answer and passage reranking, showing that a \emph{transfer-then-adapt} schedule---fine-tuning on a large general ranking corpus before adapting to the target domain---is what makes cross-encoder rankers both accurate and stable. H2CE follows this recipe by adapting a general-purpose BGE reranker to local search, and Table~\ref{tab:component_contrib} isolates the resulting gain. Because such rerankers are expensive, a complementary line of work reshapes their inference budget: Matsubara et al.~\cite{matsubara2020reranking} place a cheap ranker before the accurate transformer so that the expensive model scores only a small candidate subset, and the Cascade Transformer~\cite{soldaini2020cascade} prunes candidates progressively across the layers of a single encoder while sharing partial encodings between stages.} Poly-encoders trade quality for efficiency through partial interaction~\cite{humeau2020poly}. \rev{H2CE inherits this staged view of the reranking budget but spends it in the opposite direction: instead of using later stages only to discard candidates more cheaply, it invests additional computation in the surviving top-$K$, where the residual errors are relative rather than absolute.} Large language models can also perform pairwise or listwise reranking, but applying them to hundreds or thousands of POI candidates is difficult under real-time latency and cost constraints, especially when each candidate contains multiple fields. We therefore treat LLM reranking as a strong but expensive baseline, not as a scalable replacement for production reranking.

Standard cross-encoders are designed primarily for text, whereas POI relevance also depends on distance, rating, and review count. Raw values such as \texttt{0.3 mi}, \texttt{4.7 stars}, or \texttt{823 reviews} are poor semantic objects for attention and do not provide reliable numerical comparison. Prior work has shown that injecting retrieval scores as text can improve BERT-based reranking~\cite{askari2023bm25score,askari2024retrieverscore}, and feature-transformation work shows that neural rankers benefit from appropriate numerical preprocessing~\cite{zhuang2020feature}. \rev{A closely related line augments a transformer reranker with evidence beyond the target item. Lauriola and Moschitti~\cite{lauriola2021context} encode \emph{local} (paragraph) context in the transformer input and, to our knowledge, were among the first to supply \emph{global} context as a real-valued feature vector fused with the \texttt{[CLS]} embedding before the classification layer, explicitly identifying learning-to-rank features as the natural extension; Han et al.~\cite{han2021latency} retain much of this benefit under a serving latency budget. H2CE follows that direction with genuinely numerical attributes: distance, rating, and review count are textualized as disambiguated descriptors so that self-attention can condition on them, while exact magnitudes reach the classifier through dedicated numerical encoders fused with \texttt{[CLS]}. H2CE thus extends the score-as-text idea} from a single retrieval score to heterogeneous POI attributes: numerical values are textualized as disambiguated descriptors for cross-encoder attention while exact values are preserved through dedicated numerical encoders.
 
\paragraph{Learning to rank with structured and multi-field evidence.}
Feature-based LTR systems, including XGBoost~\cite{xgboost} and LambdaMART-style rankers~\cite{burges2010ranknet}, remain strong for structured numerical features, and tree-based models can remain competitive with neural rankers on tabular signals. Neural ranking has also been extended to multi-field and heterogeneous document representations~\cite{zamani2018multifield}. However, these approaches typically treat structured features as globally weighted ranking signals, concatenate raw values, or combine modality-specific scores through late fusion. H2CE instead uses dual numerical representation and latent-space aggregation: bucketized descriptors allow semantic--numeric attention, exact scalar MLPs preserve magnitude information, and nonlinear fusion captures conditional interactions such as distance mattering more for \emph{near me} queries and semantic match overriding proximity when a closer POI only partially satisfies the intent.
 
\paragraph{Pointwise, pairwise, and listwise ranking.}
Learning-to-rank methods are commonly categorized as pointwise, pairwise, or listwise~\cite{liu2009learning}. Pointwise models are efficient but score each candidate independently; pairwise methods such as RankNet~\cite{burges2005ranknet} and LambdaRank/ LambdaMART~\cite{burges2010ranknet} model relative preferences; listwise methods~\cite{cao2007listwise,xia2008listwise,pang2020setrank} capture richer list structure but are more expensive. Recent LLM-based ranking methods also use pairwise or listwise prompting~\cite{qin2024lrl}, and LLM-Blender~\cite{llm-blender-2023} aggregates pairwise preferences through voting. \rev{In answer selection, Zhang et al.~\cite{zhang2021joint} show that jointly encoding a target candidate together with the other top-$k$ candidates improves verification of which candidate is correct, evidence that the decisive information at the top of a ranking is relative rather than absolute---precisely the regime that H2CE Stage~2 targets.}

H2CE adopts relative comparison but specializes it to serving-constrained POI reranking. Full pairwise or listwise reranking over all candidates is impractical when the reranker may receive hundreds or thousands of POIs. H2CE separates scalability from fine-grained comparison: Stage~1 scores all candidates pointwise, while Stage~2 compares only the top-$K$ candidates where ordering errors matter most. This keeps inference cost proportional to $O(N+K(K-1))$ while preserving explicit comparison of semantic evidence, distance, rating, and review count.
 
\section{Problem Formulation}
\label{sec:problem}

\rev{This section formalizes POI reranking as a heterogeneous ranking problem over semantic, geospatial, and numerical attributes, and derives the pointwise, pairwise, and two-stage formulations used by H2CE.}

\subsection{POI Reranking}
Given a user query $q$ and a set of candidate POIs $\mathcal{D}=\{d_1,\dots,d_N\}$ produced by an upstream retriever, the reranker outputs a permutation $\pi$ of $\mathcal{D}$ that maximizes a position-weighted quality metric. Each POI $d_i$ carries two types of attributes:
 
 \textbf{Semantic attributes} $\mathbf{s}_i$: POI name, categories (\eg, \emph{restaurant, pizza}), and street address---textual fields that require language understanding to match against query intent.

 \textbf{Geospatial and numerical attributes} $\mathbf{n}_i=(z_i, r_i, c_i)$: user- or target-to-POI distance in miles, average star rating on a 0--5 scale, and review count---quantitative fields that require comparison and magnitude reasoning.

Table~\ref{tab:poi_reranking_example} illustrates the central challenge. \rev{For the
query \emph{``true Italian restaurants near the Eiffel Tower,''} a naive
distance-based reranker promotes \emph{Le Bistro Parisien} and
\emph{Tapas \& Pasta}, which are closest to the target location but do not, or
only partially, satisfy the cuisine intent; conversely, a purely semantic
reranker over-promotes \emph{Osteria Firenze}, which strongly matches the
requested cuisine but is substantially farther away. The correct ranking depends
on the interaction between semantic match and geospatial distance---a gap of a
few meters may be negligible, while a gap of more than a kilometer may reverse
the preference, especially when the query implies walking distance---and rating
and review count introduce further tradeoffs, since they provide quality and
reliability evidence but should only matter among candidates that sufficiently
match the query intent.} POI reranking therefore requires a formulation that
\rev{models} both semantic relevance and query-conditioned geo-numerical tradeoffs.

\subsection{Pointwise Formulation}
A pointwise ranker learns a scoring function $f_\theta:(q, d_i)\rightarrow \mathbb{R}$ that predicts the relevance of each candidate independently. The model is trained as a binary classifier---predicting whether $d_i$ is relevant to $q$---using binary cross-entropy loss:
\begin{equation}
    \mathcal{L}_{\text{point}} = -\big[a_i \log \sigma(f_\theta(q,d_i)) + (1-a_i)\log(1-\sigma(f_\theta(q,d_i)))\big],
\end{equation}
where $a_i \in \{0,1\}$ is the ground-truth binary relevance label. At inference, the final ranking is
\begin{equation}
    \pi=\arg\!\mathrm{sort}_{d_i\in\mathcal{D}}\, f_\theta(q,d_i),
\end{equation}
where candidates are sorted in decreasing score order.

\paragraph{Limitation.}
Pointwise scoring is efficient and can learn strong absolute relevance signals. However, because $f_\theta$ scores each POI in isolation, it does not condition a candidate's score on the attributes of the other candidates in the same result set. When multiple relevant POIs exist (\eg, several nearby Italian restaurants or three branches of the same chain), the model can separate relevant from irrelevant candidates, but it has no explicit mechanism to compare which relevant candidate is closer, better reviewed, or semantically more faithful to the query relative to the alternatives.

\subsection{Pairwise Formulation}
\label{sec:pairwise-formulation}

A pairwise ranker learns a \emph{comparison} function \[g_\phi:(q, d_i, d_j)\rightarrow [0,1],\] which models the probability $\mathbb{P}(d_i \succeq d_j \mid q)$ that $d_i$ should be ranked above $d_j$ for query $q$. Unlike pointwise scoring, the pairwise model receives \emph{both} candidates simultaneously, enabling direct comparison of their semantic and numerical attributes: which candidate better matches the query, which is closer to the target location, which has stronger review evidence, and which tradeoff is preferable for the given intent.

Let $\rho_i$ and $\rho_j$ denote the ground-truth rank positions of $d_i$ and $d_j$ within a reference ranking, where lower values indicate better ranks. In our experiments, the reference ranking is derived from a Golden Standard List produced by a leading commercial local search provider, validated \rev{with an LLM-as-a-Judge (LLMaaJ) procedure~\cite{kambhatla2026automatic}, in the spirit of earlier transformer-based automatic evaluation of ranked answers~\cite{vu2021ava}}. The binary pairwise target is
\begin{equation}
    y_{i,j}=\begin{cases}
    1 & \text{if } \rho_i \leq \rho_j,\\
    0 & \text{otherwise}.
    \end{cases}
\end{equation}
The model is trained with binary cross-entropy over these pairwise labels:
\begin{equation}
    \mathcal{L}_{\text{pair}} =
    -\big[y_{i,j}\log g_\phi(q,d_i,d_j)
    +(1-y_{i,j})\log(1-g_\phi(q,d_i,d_j))\big].
\end{equation}

\paragraph{Pairwise scoring matrix.}
Applying $g_\phi$ to all ordered non-self pairs among $K$ candidates yields a matrix $M\in\mathbb{R}^{K\times K}$ with
\[
M_{i,j}=g_\phi(q,d_i,d_j), \quad i\neq j.
\]
The diagonal entries are ignored. In the ideal case, opposite directions are probabilistically complementary:
\[
M_{i,j}+M_{j,i}=1,
\]
meaning that confidence in $d_i \succ d_j$ equals one minus confidence in $d_j \succ d_i$.

\paragraph{Copeland aggregation.}
We compute a final score for each candidate~\cite{llm-blender-2023}:
\begin{equation}
\label{pairwise}
    S_i=\sum_{\substack{j=1,j\neq i}}^{K}\big(M_{i,j}-M_{j,i}\big),
\end{equation}
and rank candidates by decreasing $S_i$. Intuitively, $S_i$ measures the net pairwise preference strength for candidate $d_i$: a POI that consistently wins head-to-head comparisons receives a high aggregate score, producing a robust total ordering even when individual comparisons contain noise.

\begin{figure*}[!ht]
  \centering
  \vspace{-1em}
  \includegraphics[width=\textwidth]{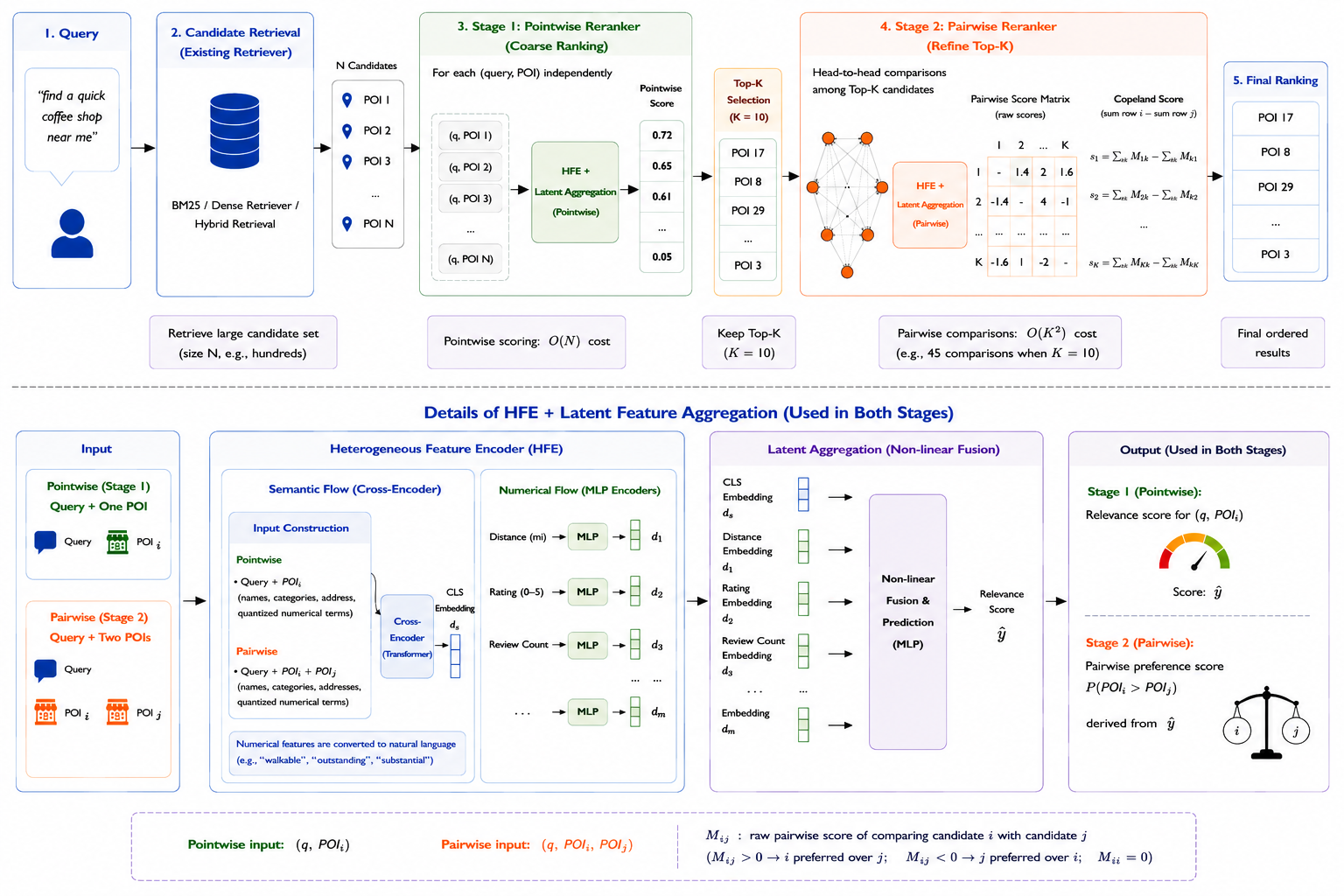}
  \Description{Diagram of the H2CE architecture. The Heterogeneous Feature Encoder splits into a semantic flow, which encodes the query, POI text, and quantized numerical descriptors with a cross-encoder, and a numerical flow, which encodes exact distance, rating, and review-count values with dedicated MLPs. The two representations are fused via latent-space aggregation. Stage 1 scores all N candidates pointwise; Stage 2 reranks the top-K candidates using K(K-1) pairwise comparisons aggregated with Copeland voting.}
  \caption{H2CE architecture. The Heterogeneous Feature Encoder uses a semantic flow that encodes the query, POI text, and quantized numerical descriptors with a cross-encoder, and a numerical flow that encodes exact distance, rating, and review-count values with MLPs. Latent aggregation fuses both representations. Stage~1 scores all $N$ candidates pointwise; Stage~2 reranks the top-$K$ using $K(K-1)$ pairwise comparisons and Copeland aggregation.}
  \label{fig:overview}
\end{figure*}

\subsection{Two-Stage Pointwise+Pairwise Ranking}
Full pairwise ranking over all $N$ candidates costs $O(N^2)$ forward passes, which is infeasible for real-time serving. For example, if $N=100$, full ordered non-self pairwise reranking requires $N(N-1)=9{,}900$ comparisons. We therefore adopt a two-stage scheme that reduces complexity to $O(N+K^2)$:
\begin{enumerate}
    \item \textbf{Stage~1 (Pointwise):} Apply $f_\theta$ to all $N$ candidates. Cost: $O(N)$.
    \item \textbf{Stage~2 (Pairwise):} Select the top-$K$ candidates by pointwise score. Apply $g_\phi$ to all $K(K-1)$ ordered non-self pairs among them. Aggregate the pairwise scores via Copeland voting. Cost: $O(K^2)$.
\end{enumerate}

The final output concatenates the Stage~2 reranked top-$K$ with the remaining candidates in Stage~1 order. With $K=10$, the pairwise stage requires $K(K-1)=90$ ordered comparisons per query, which fits within typical serving latency budgets (Section~\ref{sec:exp-runtime}).

\section{Methodology}
\label{sec:methodology}
\rev{H2CE addresses the query-conditioned semantic--geo/numerical interaction problem formalized in Section~\ref{sec:problem}, where no single feature in isolation determines whether a very close but partially matching POI should outrank a farther but semantically stronger one.} Our solution combines two complementary mechanisms. First, the Heterogeneous Feature Encoder represents distance, rating, and review count both as semantic descriptors inside the cross-encoder input and as exact scalar values processed by dedicated numerical MLPs. This allows the model to expose coarse spatial and quality concepts to transformer attention while preserving precise numerical magnitudes. Second, the two-stage architecture applies pointwise scoring for scalable filtering and pairwise comparison only among the top-$K$ candidates, where the remaining errors are often relative tradeoffs among plausible POIs. The pairwise stage turns the difficult question of which feature combination should win into a supervised comparison problem between two candidates, enabling direct feature-by-feature comparison of semantic evidence, distance, rating, and review count under bounded inference cost. Supervision is derived from a combination of a smaller customer-labeled set and a larger LLM-annotated dataset constructed on top of search-engine candidate outputs.

\subsection{System Overview}
\label{sec:system-overview}

As illustrated in Figure~\ref{fig:overview}, H2CE is a two-stage reranker operating on top of an upstream query understanding model~\cite{shah2026omniqum} and a dense retriever~\rev{\cite{chen2024bge}}. Given a query $q$ and $N$ retrieved candidate POIs $\mathcal{D}=\{d_1,\dots,d_N\}$, the system first encodes each query--POI pair with a Heterogeneous Feature Encoder (HFE). The HFE has a semantic flow, where the query, POI textual fields, and bucketized numerical descriptors are processed by a cross-encoder, and a numerical flow, where exact distance, rating, and review-count values are processed by dedicated MLPs. A latent aggregation layer fuses the cross-encoder CLS embedding with the numerical embeddings, producing a relevance representation that can express query-conditioned interactions among lexical semantics, geospatial proximity, and quality signals.

The resulting representation is used in two stages. Stage~1 applies the HFE pointwise to all $N$ candidates, producing a scalable coarse ranking and selecting the top-$K$ candidates. Stage~2 applies a pairwise version of the same heterogeneous representation to all $K(K-1)$ ordered non-self pairs among the top-$K$ candidates. Its outputs form a pairwise scoring matrix that is aggregated with Copeland voting to produce the final top-$K$ ordering. The final ranked list concatenates the pairwise-reranked top-$K$ candidates with the remaining candidates in Stage~1 order.

\subsection{Heterogeneous Feature Encoder}
\label{sec:hfe}

The HFE encodes each query--POI pair through two parallel flows designed to address the semantic--geo/numerical interaction problem. The semantic flow exposes coarse spatial and quality concepts to the cross-encoder, allowing query terms and POI text to attend to descriptors such as \emph{walkable}, \emph{near}, \emph{excellent}, or \emph{substantial}. The numerical flow preserves the exact scalar values needed to distinguish small from large distance gaps, or marginal from substantial differences in rating and review count. Together, the two flows allow the model to learn that the importance of distance, rating, and review evidence depends on the query intent and on the semantic evidence of each POI.

\paragraph{Numerical flow.}
Raw numerical attributes $\mathbf{n}_i=(z_i,r_i,c_i)$ represent distance, rating, and review count. Distance $z_i$ is computed with respect to the user location or the resolved target location expressed in the query; rating $r_i$ is the average star rating; and review count $c_i$ measures the amount of customer feedback available for the POI. These values are processed through feature-specific converters: an Online Distance Converter normalizes raw distance, a Rating Converter processes star ratings on the 0--5 scale, and a Review Count Converter handles review volume.

The converted values are then fed to three dedicated MLPs. For the pointwise stage, each MLP receives a single scalar value for one candidate. For the pairwise stage, each MLP receives a pair of values, such as $(z_i,z_j)$ for distance, enabling the model to compare the two candidates directly. This is important for geo-semantic tradeoffs: a 10-meter difference may be negligible when one POI is semantically much stronger, while a one-kilometer difference may reverse the preference, especially for proximity-oriented queries.

\paragraph{Semantic flow.}
The semantic flow converts query text, POI metadata, and bucketized numerical descriptors into a single natural-language context processed by a transformer cross-encoder. Textual POI metadata includes fields such as name, categories, and street address. Numerical values are also converted into disambiguated natural-language descriptors, allowing them to participate in semantic matching. For example, a pointwise input may be represented as:
\begin{quote}
\small
\emph{Query: true Italian restaurants near the Eiffel Tower. POI: NAME: Roma Trattoria; CATEGORIES: Italian restaurant, pasta, Roman cuisine; DISTANCE: walkable; RATING: outstanding; REVIEWS: substantial; ADDRESS: ...}
\end{quote}

For pairwise inputs, the context extends to include both candidates:
\begin{quote}
\small
\emph{Query: true Italian restaurants near the Eiffel Tower. POI A: ... DISTANCE: very close; RATING: excellent; REVIEWS: abundant. [SEP] POI B: ... DISTANCE: walkable; RATING: outstanding; REVIEWS: substantial.}
\end{quote}

This representation allows the cross-encoder to condition semantic matching on coarse spatial and quality concepts. For example, the model can learn that \emph{near} or \emph{walkable} should matter more when the query expresses proximity, while \emph{outstanding} ratings and \emph{substantial} reviews should matter more when the query asks for the \emph{best} or most reliable option. At the same time, the exact numerical MLP pathways preserve the scalar precision that is lost when values are bucketized.

\paragraph{Semantic quantization.}
Transformers do not reliably compare raw numerical tokens such as \texttt{0.3 mi} and \texttt{2.1 mi}, or \texttt{4.6 stars} and \texttt{4.8 stars}. We therefore project numerical values into a vocabulary of disambiguated natural-language descriptors. Distance, rating, and review count use separate vocabularies to avoid cross-feature collisions. For instance, distance bins may use terms such as \emph{here}, \emph{walkable}, \emph{near}, or \emph{far}; rating bins may use terms such as \emph{good}, \emph{excellent}, or \emph{outstanding}; and review-count bins may use terms such as \emph{limited}, \emph{substantial}, or \emph{abundant}. These descriptors provide the cross-encoder with semantic hooks for numerical information, while the numerical MLPs retain exact values.

\subsection{Latent-Space Feature Aggregation}
\label{sec:latent-aggregation}

Traditional heterogeneous rankers often reduce each pathway to a scalar score and aggregate them through a weighted sum:
\[
s = \lambda_s h_s + \lambda_d h_d + \lambda_r h_r + \lambda_c h_c .
\]
This late scalar fusion is limited because it cannot easily represent conditional interactions among semantic evidence, distance, rating, and review count. For example, distance should dominate when the query asks for the \emph{closest} result and the candidates are semantically similar; semantic match should dominate when a closer POI only partially satisfies the requested intent; rating and review count should matter more when the query asks for the \emph{best} or most reliable option.

H2CE instead performs latent-space aggregation, \rev{extending the projected-context-vector fusion with \texttt{[CLS]} of Lauriola and Moschitti~\cite{lauriola2021context} to multiple learned numerical encoders.} The cross-encoder produces a CLS embedding $h_{\mathrm{cls}}\in\mathbb{R}^d$, and each numerical MLP produces a dense embedding rather than a scalar. Specifically, distance, rating, and review count are mapped to embeddings:
\[
\tilde{h}_d = E_d(\cdot), \quad
\tilde{h}_r = E_r(\cdot), \quad
\tilde{h}_c = E_c(\cdot),
\]
where each numerical encoder is a progressive MLP with GELU activations and a 256-dimensional output. The fused representation is:
\begin{equation}
    h_{\mathrm{fused}} =
    [h_{\mathrm{cls}};\tilde{h}_d;\tilde{h}_r;\tilde{h}_c]
    \in \mathbb{R}^{d+3\cdot256}.
\end{equation}
A final MLP classifier $E_{\mathrm{cls}}$ maps this fused representation to a scalar score:
\begin{equation}
    s = E_{\mathrm{cls}}(h_{\mathrm{fused}};\theta_{\mathrm{cls}}).
\end{equation}

This design allows the classifier to learn nonlinear feature interactions in representation space, \rev{such as preferring a slightly farther POI whose semantic and review evidence is stronger.} Latent aggregation is used in both stages: in Stage~1 it produces pointwise relevance scores, while in Stage~2 it produces pairwise preference scores.

\subsection{Stage 1: Pointwise HCE}
\label{sec:stage1}

The pointwise ranker scores each $(q,d_i)$ pair independently. It applies the HFE to produce four representations: a cross-encoder CLS embedding from the semantic flow, and three numerical embeddings from the distance, rating, and review-count MLPs. These embeddings are fused through latent-space aggregation to produce a pointwise relevance score:
\[
f_\theta(q,d_i) \in \mathbb{R}.
\]

Stage~1 is responsible for scalable coarse ranking. It must process all $N$ candidates returned by the upstream retriever, so it cannot perform expensive full pairwise or listwise reasoning over the entire candidate set. Its role is to identify candidates that are likely to be relevant according to semantic match, geospatial proximity, and numerical quality signals. This makes Stage~1 critical: if a semantically strong or geographically appropriate POI is not placed in the top-$K$, the pairwise stage cannot recover it.

\rev{From the geo-semantic perspective, Stage~1 learns \emph{absolute} evidence per candidate. Because each candidate is scored independently, however, it cannot resolve tradeoffs that depend on the other candidates in the set, so its output serves as a high-recall filter for Stage~2.}

\subsection{Stage 2: Pairwise HCE}
\label{sec:stage2}

\paragraph{The positive-over-positive problem.}
Pointwise ranking can learn strong absolute relevance signals, but it does not directly observe the competing candidates whose attributes determine the final tradeoff. This is limiting when several POIs are plausible and the correct decision depends on relative differences rather than an absolute relevance scale. For example, for \emph{``true Italian restaurants near the Eiffel Tower,''} the ranking decision may depend on whether a very close mixed-cuisine POI should outrank a farther but semantically stronger Italian restaurant. There is no universal absolute score for this tradeoff: the preferred candidate depends on the semantic evidence of both POIs, the distance gap between them, and the query intent.

The pairwise stage converts this into a direct supervised comparison problem. Given two candidates, the model can compare feature by feature: the distance of $d_i$ versus the distance of $d_j$, the cuisine evidence of $d_i$ versus that of $d_j$, and the rating and review evidence of one candidate versus the other. Pairwise preference labels are derived from the available supervision: a smaller customer-labeled set and a larger LLM-annotated dataset constructed over search-engine candidate outputs. Training on these pairwise preferences teaches the model which feature combinations tend to win in context. Thus, pairwise reranking partially solves the semantic--geo/numerical tradeoff by making the relative decision explicit, while restricting comparison to the top-$K$ candidates keeps the cost practical.

\paragraph{Pairwise forward pass.}
Given a query $q$ and a pair of candidates $(d_i,d_j)$, the pairwise ranker reuses the HFE with inputs adapted for comparison. In the semantic flow, the cross-encoder receives the concatenated context $[q;\mathrm{SEP};d_i;\mathrm{SEP};d_j]$, where each POI includes textual fields and bucketized descriptors for distance, rating, and review count. This enables joint attention over both candidates. The model can learn, for example, that $d_i$ has stronger cuisine evidence, while $d_j$ is closer or has stronger review evidence.

In the numerical flow, each MLP receives a pair of values: \[ (z_i,z_j), \quad (r_i,r_j), \quad (c_i,c_j), \] for distance, rating, and review count respectively. This paired representation makes numerical comparison explicit. The model can learn that a small distance gap may not compensate for weak semantic match, while a large distance gap may become decisive when both candidates satisfy the query intent. The pairwise embeddings are fused through latent-space aggregation to produce: $ g_\phi(q,d_i,d_j) \in [0,1],$ which represents the model's confidence that $d_i$ should rank above $d_j$ for query $q$.

\paragraph{Copeland aggregation.}
We evaluate every ordered non-self pair among the top-$K$ candidates, producing the pairwise scoring matrix $M$ of Section~\ref{sec:pairwise-formulation}; diagonal entries are ignored. The final score, $S_i$, for each candidate is computed with Eq.~\ref{pairwise}.
Candidates are ranked by decreasing $S_i$. This aggregation rewards candidates that consistently win head-to-head comparisons against the other top-$K$ candidates. It is particularly suitable for POI reranking because the final decision often depends on multiple relative tradeoffs rather than a single absolute relevance score.

\subsection{Inference Complexity}
\label{sec:inference-complexity}

\rev{H2CE separates scalable filtering from fine-grained comparison. Stage~1 applies pointwise scoring to all candidates at cost $O(N)$, and Stage~2 applies pairwise comparison only to the top-$K$ candidates at cost $O(K(K-1))$, i.e., $O(K^2)$ asymptotically, giving a total inference cost of}
\[
O(N + K(K-1)).
\]
\rev{With $K=10$, Stage~2 requires only $90$ ordered non-self comparisons per query, rather than the $N(N-1)$ of full pairwise reranking, which is infeasible when $N$ may be hundreds or thousands. Explicit comparison is thus applied where it is most valuable: among top candidates that are already semantically and geographically plausible.}

\begin{table*}[t]
\centering
\caption{Full comparison on the component test set (5,743 queries). Models are grouped: unsupervised/LTR baselines, zero-shot
neural models, fine-tuned pointwise variants, and two-stage systems. All values are percentages (\%). Best results in bold.}
\label{tab:baselines}
\renewcommand{\arraystretch}{1.1}
\small
\setlength{\tabcolsep}{4pt}
\begin{tabular}{@{}l|cccc|cccc|cc|c@{}}
\toprule
\multirow{2}{*}{\textbf{Model}} & \multicolumn{4}{c|}{\textbf{RDQ}} & \multicolumn{4}{c|}{\textbf{NDCG}} &
\multicolumn{2}{c|}{\textbf{Precision}} & \multirow{2}{*}{\textbf{MRR}} \\
\cmidrule(lr){2-5} \cmidrule(lr){6-9} \cmidrule(lr){10-11}
& @1 & @5 & @10 & @20 & @1 & @5 & @10 & @20 & @1 & @5 & \\
\midrule
BM25-only & 9.00 & 11.33 & 11.93 & 12.41 & 9.68 & 12.26 & 13.91 & 16.83 & 9.68 & 6.86 & 19.17 \\
XGBoost LTR & 45.07 & 39.81 & 37.87 & 37.04 & 47.88 & 44.66 & 44.84 & 49.12 & 47.88 & 35.64 & 58.97 \\
\midrule
RankGPT (Claude Haiku) & 22.19 & 22.93 & 21.70 & 20.99 & 23.80 & 31.59 & 35.91 & 39.74 & 23.80 & 16.10 & 34.89 \\
BGE Zero-shot & 22.56 & 24.08 & 23.95 & 24.07 & 24.62 & 26.76 & 28.03 & 32.03 & 24.62 & 18.78 & 38.32 \\
\midrule
BGE Fine-tuned & 57.29 & 51.36 & 48.04 & 46.33 & 61.14 & 57.86 & 56.84 & 60.08 & 61.14 & 44.96 & 70.81 \\
PW HCE w/ Scalar Agg. & 61.65 & 55.54 & 52.27 & 50.50 & 65.61 & 62.26 & 61.45 & 64.77 & 65.61 & 48.53 & 74.12 \\
Pointwise HCE (Ours) & 64.29 & 58.68 & 55.14 & 53.01 & 68.38 & 65.50 & 64.54 & 67.54 & 68.38 & 51.61 & 76.31 \\
\midrule
H2CE w/ Scalar Agg. & 66.14 & 59.64 & 54.38 & 52.34 & 69.77 & 66.38 & 63.21 & 66.40 & 69.77 & 52.12 & 77.02 \\
H2CE (Ours) & \textbf{67.02} & \textbf{60.77} & \textbf{56.16} & \textbf{53.91} & \textbf{70.64} & \textbf{67.48} & \textbf{65.27}
& \textbf{68.19} & \textbf{70.64} & \textbf{53.42} & \textbf{77.62} \\
\bottomrule
\end{tabular}
\end{table*}

\begin{table*}[t]
\centering
\caption{Component contribution analysis. Each row adds one architectural component on top of the previous, showing the progressive value of domain fine-tuning, numerical features, latent-space aggregation, and pairwise reranking. All values are percentages (\%).}
\label{tab:component_contrib}
\small
\resizebox{\textwidth}{!}{
\begin{tabular}{@{}l|cccc|cccc|cc|c|l@{}}
\toprule
\multirow{2}{*}{\textbf{Model}} & \multicolumn{4}{c|}{\textbf{RDQ}} & \multicolumn{4}{c|}{\textbf{NDCG}} & \multicolumn{2}{c|}{\textbf{Precision}} & \multirow{2}{*}{\textbf{MRR}} & \multirow{2}{*}{\textbf{Component Added}} \\
\cmidrule(lr){2-5} \cmidrule(lr){6-9} \cmidrule(lr){10-11}
 & @1 & @5 & @10 & @20 & @1 & @5 & @10 & @20 & @1 & @5 & & \\
\midrule
BGE Zero-shot & 22.56 & 24.08 & 23.95 & 24.07 & 24.62 & 26.76 & 28.03 & 32.03 & 24.62 & 18.78 & 38.32 & Pretrained cross-encoder \\
BGE Fine-tuned & 57.29 & 51.36 & 48.04 & 46.33 & 61.14 & 57.86 & 56.84 & 60.08 & 61.14 & 44.96 & 70.81 & + Domain fine-tuning \\
PW HCE w/ Scalar Agg. & 61.65 & 55.54 & 52.27 & 50.50 & 65.61 & 62.26 & 61.45 & 64.77 & 65.61 & 48.53 & 74.12 & + Numerical MLP heads \\
Pointwise HCE (Ours) & 64.29 & 58.68 & 55.14 & 53.01 & 68.38 & 65.50 & 64.54 & 67.54 & 68.38 & 51.61 & 76.31 & + Latent-space aggregation \\
H2CE (Ours) & \textbf{67.02} & \textbf{60.77} & \textbf{56.16} & \textbf{53.91} & \textbf{70.64} & \textbf{67.48} & \textbf{65.27} & \textbf{68.19} & \textbf{70.64} & \textbf{53.42} & \textbf{77.62} & + Pairwise reranking \\
\bottomrule
\end{tabular}}
 
\end{table*}

\section{Experiments}
\label{sec:exp}
\rev{We evaluate H2CE on the component test set (5,743 queries), first against external baselines and then by ablating each design decision in turn: numerical features across both stages, the pointwise training strategy, the pairwise stage and its sensitivity to $K$, inference latency, a half-matrix optimization, and throughput.}

\subsection{Experimental Setup}
\label{sec:exp-setup}
\paragraph{Training data.}
We train on roughly 500k queries with query--POIs data. In our experiments, the reference ranking is derived from a Golden Standard List (GSL) produced by a leading commercial local search provider, validated by LLMaaJ, which is used to train both pointwise reranker and pair-wise reranker models~\cite{kambhatla2026automatic}. Specifically, Positive POIs are those appearing in the GSL; negatives are candidates outside it. For the pairwise stage, we additionally condition on the Stage~1 pointwise scores and sample from the top-10 candidates, aligning the training distribution with the inference setting where Stage~2 compares only highly ranked candidates.

\paragraph{Test set.}
We evaluate on a component test set of 5,743 queries from an enriched pointwise local-search dataset, isolating reranker performance from upstream routing effects. Each query has a candidate pool containing positive and negative POIs, with ground-truth rankings derived from the corresponding GSL.

\paragraph{Metrics.}
Our primary metric is Rank-Deviation Quality (RDQ@$k$), a position-weighted correctness metric \rev{for ranking settings in which} a query may have many valid answers and quality depends on their relative ordering. Unlike metrics that require absolute relevance grades, RDQ evaluates a system ranking against an ordinal gold-standard list and penalizes deviations from the target order. This makes it suitable for multi-answer retrieval and reranking scenarios, such as points of interest, products, entities, or other candidate lists where many items may be valid but should still be ordered by preference~\cite{zhou2026rankdeviation}. We also report Normalized Discounted Cumulative Gain (NDCG@$k$), Precision@$k$, and Mean Reciprocal Rank (MRR).

\paragraph{Baselines.}
\textbf{BM25-only} ranks candidates by descending BM25 score, with no learned reranking, establishing the lexical retrieval-quality floor. \textbf{XGBoost LTR} is an XGBRanker~\cite{xgboost} trained with the \texttt{rank:ndcg} objective (LambdaMART) on same training queries using five features: geographic distance (54.9\% importance), BM25 score (17.1\%), name n-gram Jaccard similarity (16.4\%), review count (8.9\%), and rating (2.7\%). This represents the class of efficient feature-based rankers commonly deployed in commercial local search systems. \textbf{RankGPT (Claude Haiku)} uses Claude~3 Haiku~\cite{anthropic2024claude3} with the RankGPT~\cite{sun2023chatgpt} sliding-window listwise permutation method (window size 20, step size 10). Each POI is represented by name, categories, distance, address, and rating, and review count. This baseline evaluates zero-shot LLM reranking quality under a bounded candidate setting, but is not intended as a scalable production solution over hundreds or thousands of candidates. \textbf{BGE Zero-shot} applies BGE-reranker-large~\cite{bge_embedding} as a pretrained cross-encoder with no local-search fine-tuning, scoring each $(q,d_i)$ pair via the raw CLS logit. \textbf{Pointwise HCE w/ Scalar Aggregation} combines a cross-encoder, numerical quantization, and numerical MLPs using scalar weighted-sum fusion. \textbf{Pointwise HCE} is the full Stage~1 ranker with cross-encoder input quantization, numerical MLPs, and latent-space aggregation, trained with the multi-pair strategy. \textbf{H2CE (Ours)} is the full two-stage system: Pointwise HCE Stage~1 followed by a Stage~2 pairwise reranker with latent-space aggregation, Copeland voting, and top-$K$ aligned training.

\paragraph{Implementation.}
 
We use BGE-reranker-large~\cite{bge_embedding} as the cross-encoder backbone for both stages. All stages train with binary cross-entropy at learning rate $10^{-5}$. The pointwise model trains with up to 16 positive and 16 negative POIs per query. Pairwise training samples $K{=}4$ POIs per query from the Stage~1 top-10 and constructs ordered comparison pairs from this subset. All training \rev{runs are} conducted on 8$\times$ A100-40GB GPUs.

\subsection{Component-Level Comparison}
 
\label{sec:exp-component}
Table~\ref{tab:baselines} compares all models on the 5,743-query component test set, isolating reranker performance from upstream routing effects. BM25-only reaches only 9.00\% RDQ@1, confirming that lexical matching alone is weak for local search. The zero-shot neural models, RankGPT and BGE Zero-shot, perform similarly at approximately 22\% RDQ@1, showing that generic LLM/listwise reranking and pretrained cross-encoder matching remain far below supervised local-search rerankers without domain adaptation. XGBoost LTR reaches 45.07\% RDQ@1 using only five structured features, highlighting the importance of explicit numerical and lexical signals, especially geographic distance.

The neural results show cumulative gains from each component. Domain fine-tuning improves BGE by +34.73\% RDQ@1 over zero-shot; adding numerical MLPs contributes +4.36\%; replacing scalar fusion with latent aggregation adds +2.64\%; and the pairwise stage adds +2.73\%. The full H2CE system achieves 67.02\% RDQ@1, improving over XGBoost LTR by +21.95\% absolute and over zero-shot BGE by +44.46\%. Table~\ref{tab:component_contrib} isolates these increments.

\subsection{Numerical Feature Ablation Across Stages}
\label{sec:exp-hfe-ablation}

Table~\ref{tab:numerical_ablation} ablates numerical MLP inputs independently in the pointwise and pairwise stages, showing where distance, rating, and review count matter.
\begin{table}[!ht]
\centering
\caption{Numerical feature ablation across stages. \rev{``Stage~1 Num.'' and ``Stage~2 Num.'' indicate whether the numerical MLP inputs are enabled in the pointwise and pairwise stages, respectively.}}
\label{tab:numerical_ablation}
\small
\setlength{\tabcolsep}{3.5pt}
\resizebox{\columnwidth}{!}{
\begin{tabular}{@{}cc|cccc@{}}
\toprule
\rev{\textbf{Stage 1 Num.}} & \rev{\textbf{Stage 2 Num.}} & \textbf{P@1} & \textbf{NDCG@5} & \textbf{RDQ@1} & \textbf{RDQ@5} \\
\midrule
\ding{55} & \ding{55} & 55.49 & 50.13 & 51.65 & 44.06 \\
\ding{51} & \ding{55} & 66.86 & 65.46 & 62.64 & 58.66 \\
\ding{55} & \ding{51} & 61.95 & 53.87 & 57.81 & 47.07 \\
\ding{51} & \ding{51} & \textbf{70.64} & \textbf{67.48} & \textbf{67.02} & \textbf{60.77} \\
\bottomrule
\end{tabular}}
\end{table}
The semantic-only model still reaches 51.65\% RDQ@1, showing that textual matching is substantial but insufficient. Adding numericals only to Stage~1 gives a larger gain (+10.99\% RDQ@1) than adding them only to Stage~2 (+6.16\%), because Stage~1 determines which candidates enter the top-$K$. The full model performs best, confirming that numerical features are useful both for coarse filtering and for fine-grained pairwise comparison.

\subsection{Training Strategy}
\label{sec:exp-training}
\textit{Pointwise: multi-pair vs.\ single-pair.}
Table~\ref{tab:training_strategy} compares single-pair training, with one positive and one negative POI per query, against multi-pair training, which samples up to 16 positives and 16 negatives.

\begin{table}[!ht]
\centering
\caption{Pointwise training strategy ablation.}
\label{tab:training_strategy}
\small
\setlength{\tabcolsep}{3.5pt}
\resizebox{\columnwidth}{!}{
\begin{tabular}{@{}l|cccc|c@{}}
\toprule
\textbf{Training Strategy} & \textbf{RDQ@1} & \textbf{RDQ@5} & \textbf{P@1} & \textbf{P@5} & \textbf{MRR} \\
\midrule
Single-pair (1 pos, 1 neg) & 61.65 & 55.54 & 65.61 & 48.53 & 74.12 \\
Multi-pair (up to 16 pos/neg) & \textbf{64.29} & \textbf{58.68} & \textbf{68.38} & \textbf{51.61} & \textbf{76.31} \\
\midrule
$\Delta$ [bps] & +264 & +314 & +277 & +308 & +219 \\
\bottomrule
\end{tabular}}
\end{table}
Multi-pair training improves RDQ@1 by +2.64\% and RDQ@5 by +3.14\%, likely because the model observes richer within-query contrasts and lower-variance gradients.

\textit{Pairwise: top-$K$ aligned sampling.}
At inference, Stage~2 compares only the top-$K$ Stage~1 candidates, a pool biased toward plausible POIs with subtle differences. Uniform sampling from the full candidate set would train on an easier distribution. We therefore select the Stage~1 top-10 candidates and randomly sample 4 of them, forming 12 ordered comparisons per query. This aligns training with the inference regime and lets pair composition vary with query difficulty.

\subsection{Does the Pairwise Stage Help?}
\label{sec:exp-pairwise}

Table~\ref{tab:pairwise} adds the pairwise stage on top of Pointwise HCE with $K{=}10$. Pairwise reranking improves all metrics, with the largest gains at the top of the list: +2.73\% RDQ@1, +2.09\% RDQ@5, +2.26\% NDCG@1, and +1.98\% NDCG@5. This matches the intended role of Stage~2: selecting the best result among multiple plausible candidates. Deeper metrics improve less because only the top-10 order is changed.

\begin{table}[!ht]
\centering
\caption{Pairwise stage ablation ($K{=}10$).}
\label{tab:pairwise}
\small
\setlength{\tabcolsep}{3.5pt}
\resizebox{\columnwidth}{!}{
\begin{tabular}{@{}l|ccc|ccc|c@{}}
\toprule
\multirow{2}{*}{\textbf{Model}} & \multicolumn{3}{c|}{\textbf{RDQ}} & \multicolumn{3}{c|}{\textbf{NDCG}} & \multirow{2}{*}{\textbf{MRR}} \\
\cmidrule(lr){2-4} \cmidrule(lr){5-7}
 & @1 & @5 & @10 & @1 & @5 & @10 & \\
\midrule
Pointwise HCE & 64.29 & 58.68 & 55.14 & 68.38 & 65.50 & 64.54 & 76.31 \\
H2CE  & \textbf{67.02} & \textbf{60.77} & \textbf{56.16} & \textbf{70.64} & \textbf{67.48} & \textbf{65.27} & \textbf{77.62} \\
\midrule
$\Delta$ [bps] & +273 & +209 & +102 & +226 & +198 & +73 & +131 \\
\bottomrule
\end{tabular}}
\end{table}

\begin{table}[!ht]
\centering
\caption{Top-$K$ sensitivity of the pairwise stage.}
\label{tab:topk}
\small
\setlength{\tabcolsep}{3.5pt}
\resizebox{\columnwidth}{!}{
\begin{tabular}{@{}c|c|cccc|c|c@{}}
\toprule
\textbf{$K$} & \textbf{Pairs/q} & \textbf{RDQ@1} & \textbf{RDQ@5} & \textbf{P@1} & \textbf{NDCG@10} & \textbf{MRR} & \textbf{ms/q} \\
\midrule
-- & 0 & 64.29 & 58.68 & 68.38 & 64.54 & 76.31 & 16.5 \\
\midrule
5  & 20  & 66.84 & 59.16 & 70.47 & 64.97 & 77.46 & 23.5 \\
10 & 90  & \textbf{67.02} & \textbf{60.77} & \textbf{70.64} & 65.27 & \textbf{77.62} & 33.1 \\
20 & 380 & 66.55 & 60.28 & 70.22 & \textbf{65.87} & 77.31 & 77.0 \\
30 & 870 & 65.76 & 59.81 & 69.44 & 65.39 & 76.87 & 152.6 \\
40 & 1560 & 65.53 & 59.64 & 69.21 & 65.16 & 76.71 & 260.3 \\
\bottomrule
\end{tabular}}
\end{table}

\begin{table}[t]
\centering
\caption{Per-query inference latency (single A100-40GB).}
\label{tab:latency}
\footnotesize
\setlength{\tabcolsep}{3pt}
\resizebox{0.85\columnwidth}{!}{
\begin{tabular}{@{}l|c|cccc@{}}
\toprule
\textbf{Model} & \textbf{Pairs} & \textbf{Mean} & \textbf{Median} & \textbf{P90} & \textbf{P99} \\
\midrule
BGE Zero-shot & 0 & 74.4 & 64.3 & 110.4 & 153.3 \\
BGE Fine-tuned & 0 & 74.4 & 64.4 & 110.5 & 153.3 \\
Pointwise HCE & 0 & 74.6 & 64.6 & 110.7 & 153.7 \\
\midrule
H2CE (Full matrix) & 90 & 171.5 & 163.2 & 216.9 & 263.5 \\
H2CE (Half-matrix) & 45 & 127.8 & 120.1 & 171.7 & 212.3 \\
\bottomrule
\end{tabular}}
\end{table}

\subsection{Top-$K$ Sensitivity}
\label{sec:exp-topk}

Table~\ref{tab:topk} shows the quality--cost tradeoff for the pairwise top-$K$. At $K{=}5$, Stage~2 is recall-limited by Stage~1. Performance peaks at $K{=}10$ for RDQ@1 and P@1, \rev{so we use $K{=}10$ throughout}. Larger values marginally help NDCG@10 but degrade top-biased metrics, likely because inference diverges from the top-10 training distribution while cost grows quadratically. The ``ms per query'' reports time from batched multi-GPU evaluation (total wall-clock divided by 5,743 queries on 8$\times$A100), which benefits from GPU parallelism and batching and differs from the following latency analysis.

\subsection{Latency Analysis}
\label{sec:exp-runtime}

We measure per-query latency on a single A100-40GB GPU with batch size 1, 100 measured queries, and 10 warmup queries. Table~\ref{tab:latency} reports latency percentiles. The pointwise models share the same cross-encoder backbone and achieve approximately 65\,ms median latency; numerical MLPs add negligible overhead. Full H2CE with 90 pairwise comparisons increases median latency to 163\,ms, while the half-matrix variant reduces it to 120\,ms.

\subsection{Half-Matrix Optimization}
\label{sec:exp-halfmatrix}

For $K{=}10$, full pairwise scoring evaluates $K(K-1)=90$ ordered non-self pairs. The half-matrix variant scores only $\binom{K}{2}=45$ unordered pairs and reconstructs the opposite direction using probabilistic complementarity, $M_{j,i}=1-M_{i,j}$. Table~\ref{tab:half_matrix} shows that this loses less than 0.3\% relative on all metrics while reducing median latency by 26.4\% (163\,ms to 120\,ms), making it attractive for tight serving budgets.

\subsection{Runtime TPS}
\label{sec:exp-throughput}

Table~\ref{tab:throughput} estimates serving throughput from median single-GPU latency. The full H2CE system reaches approximately 6 TPS on one A100 GPU, or 49 TPS on an 8-GPU node; the half-matrix variant improves this to 8.3 TPS per GPU, or 67 TPS on 8 GPUs, with negligible quality loss. Pointwise HCE is faster at 15.5 TPS per GPU, while RankGPT is limited to approximately 9 TPS and is not scalable to large candidate pools. These results show that H2CE provides pairwise reranking quality at practical serving cost.

\begin{table}[t]
\centering
\caption{Half-matrix pairwise ablation.}
\label{tab:half_matrix}
\small
\setlength{\tabcolsep}{3pt}
\resizebox{\columnwidth}{!}{
\begin{tabular}{@{}l|c|ccc|c|c@{}}
\toprule
\textbf{Variant} & \textbf{Pairs} & \textbf{RDQ@1} & \textbf{RDQ@5} & \textbf{P@1} & \textbf{MRR} & \textbf{Median ms} \\
\midrule
H2CE (Full matrix) & 90 & \textbf{67.02} & \textbf{60.77} & \textbf{70.64} & \textbf{77.62} & 163.2 \\
H2CE (Half-matrix) & 45 & 66.85 & 60.72 & 70.49 & 77.53 & 120.1 \\
\midrule
$\Delta$ (rel.) & $-$50\% & $-$0.25\% & $-$0.08\% & $-$0.21\% & $-$0.12\% & \textbf{$-$26.4\%} \\
\bottomrule
\end{tabular}}
\end{table}

\begin{table}[t]
\centering
\caption{Estimated serving throughput from single-GPU median latency.}
\label{tab:throughput}
\small
\setlength{\tabcolsep}{3pt}
\resizebox{\columnwidth}{!}{
\begin{tabular}{@{}l|c|c|c@{}}
\toprule
\textbf{Model} & \textbf{Median latency} & \textbf{TPS/GPU} & \textbf{TPS/8 GPUs} \\
\midrule
BM25 (CPU) & $<$1\,ms & $>$1000 & -- \\
XGBoost LTR (CPU) & $\sim$5\,ms & $\sim$200 & -- \\
RankGPT (Claude Haiku, API) & 111\,ms & 9.0 & -- \\
\midrule
Pointwise HCE & 64.6\,ms & 15.5 & 124 \\
H2CE ($K{=}10$, full) & 163.2\,ms & 6.1 & 49 \\
H2CE ($K{=}10$, half-matrix) & 120.1\,ms & 8.3 & 67 \\
\bottomrule
\end{tabular}}
\end{table}
 
\section{Conclusion}
\label{sec:conclusion}

We presented H2CE, a serving-constrained POI reranker designed to model query-conditioned semantic--geo/numerical interactions. H2CE combines two components: a \emph{Heterogeneous Feature Encoder with latent-space aggregation}, which exposes distance, rating, and review count both as bucketized semantic descriptors inside the cross-encoder and as exact scalar values through numerical MLPs; and a \emph{two-stage pointwise-plus-pairwise architecture}, which uses pointwise scoring for scalable filtering and pairwise Copeland-aggregated comparison among the top-$K$ candidates. \rev{This reduces pairwise cost from $O(N^2)$ to $O(N+K(K-1))$.} On a 5,743-query local search test set, H2CE achieves 67.02\% RDQ@1, improving over XGBoost LTR by +21.95\% absolute and over a zero-shot LLM reranker by +44.83\%. The pairwise stage contributes +2.73\% RDQ@1 over the pointwise model alone, confirming the value of explicit head-to-head comparison among highly plausible POIs.
 
\paragraph{Limitations.}
\rev{H2CE improves ranking quality at the cost of latency: even with the half-matrix optimization, median latency rises from 65\,ms for pointwise-only reranking to 120\,ms for the two-stage system, so the pointwise model may be preferable under very tight budgets.} A second limitation is supervision quality. Our training data combines a smaller customer-labeled set with a larger LLM-annotated dataset constructed over search-engine candidate outputs. \rev{These labels may contain duplicated POIs, ambiguous preferences, or imperfect judgments, which can cap measured performance; manual inspection indicates that some apparent H2CE errors are in fact reasonable rankings penalized by noisy labels.}
 
\paragraph{Reproducibility.}
Although our dataset is proprietary, \rev{we release POISS~\cite{maritan2026poiss}, a large-scale multilingual POI search dataset built with the same methodology plus privacy-preserving techniques enabling open-source access. It provides comparable query--POI sets and annotations, and should yield comparable, though not identical, results. All architecture and training details are specified above, and the BGE-reranker-large backbone~\cite{bge_embedding} is publicly available.}

\section*{GenAI Usage Disclosure}
In compliance with ACM policy, we disclose all uses of generative AI tools in this work:

Claude was utilized to generate sections of this Work, including text, graphs and formatting assistance during manuscript preparation; all scientific content, experimental design, and results were produced by the authors. Claude 3 Haiku was additionally used as a zero-shot LLM reranker baseline (RankGPT prompting), reported as an experimental baseline in Section~\ref{sec:exp}.




\bibliographystyle{ACM-Reference-Format}
\balance
\bibliography{references}
\end{document}